\documentclass[12pt,a4paper]{amsart}
\usepackage[french]{babel}
\usepackage[latin1]{inputenc}
\usepackage[T1]{fontenc}
\usepackage{mathrsfs,dsfont}
\usepackage{amssymb,amsmath}
\usepackage{url}

\def\pacs#1{\textbf{P.A.C.S.:}\ #1}

\newcommand{\ep}{\varepsilon}

\begin{document}
\title{Introduction à l'\oe{}uvre de S. Kichenassamy en Physique Th\'eorique}
\author{Satyanad Kichenassamy}
\address{Laboratoire de Math\'ematiques, Universit\'e de Reims Champagne-Ardenne, Moulin de la Housse, B.P. 1039, F-51687 Reims Cedex 2, France.}

\email{satyanad.kichenassamy@univ-reims.fr}%

\date{12 ao\^ut 2016 (Aug.~12, 2016)}

\vskip 1em

\hfill%
\emph{Annales de la Fondation Louis de Broglie} {\bf 41}  (2016), 131--151.

\hfill%
\url{http://aflb.ensmp.fr/AFLB-411/aflb411m862.pdf}

\vskip 1em

\maketitle

\markboth{L'\oe{}uvre de S. Kichenassamy en Physique Th\'eorique}{L'\oe{}uvre de S. Kichenassamy en Physique Th\'eorique}

\vskip 1cm
\begin{abstract}
L'\oe uvre de S. Kichenassamy (1926--2015) en Physique Théorique et Math\'ematique couvre l'ensemble du spectre de la Physique Relativiste, ainsi que certaines théories unitaires : de la clarification des postulats sur lesquels se fondent les deux théories de la Relativité, aux domaines d'application tels que la mesure du temps propre, la formation des images, la théorie des collisions, les théories cinétiques et le transfert de rayonnement, ou les pulsars. Après quelques repères biographiques, on montre, en reprenant un raisonnement formulé en 1963, et développé dans une série de travaux, que l'interprétation physique de la Relativité Générale requiert d'abandonner le principe d'équivalence forte au profit de la \emph{C-équivalence}. Cette analyse, en précisant la notion mathématique d'observateur, conduit également à des résultats nouveaux en Relativité Restreinte. Les applications sont résumées. Ses contributions à l'Indologie seront développées dans un autre travail.

{\it 
\textsc{English abstract.}  The work of S. Kichenassamy (1926--2015) in Theoretical and Mathematical Physics covers the spectrum of Relativistic Physics: from the clarification of the postulational basis of the two theories of Relativity, to applications to the measurement of proper time, image formation, collision theory, kinetic theory and radiative transfer, or pulsar electrodynamics, to name a few. After brief biographic remarks, we show, by following an argument first formulated in 1963, and developed in a series of papers, that the physical interpretation of General Relativity requires the replacement of the strong equivalence principle by the \emph{C-equivalence principle}. By giving a mathematical status to the observer, his insight yields new results regarding measurements actual observers can make, both in Special and in General Relativity. Applications are outlined. His contributions to Indology will be discussed elsewhere.}
\end{abstract}


\vskip 2em

\section{Introduction}

L'\oe uvre de S. Kichenassamy\footnote{Le père de l'auteur.} (1926--2015), Directeur de Recherche au C.N.R.S., s'est développée principalement en Physique Théorique et en Physique Math\'ematique. Ses contributions couvrent tout le champ de la Physique Relativiste. Son impact sur l'Indologie \cite{ind-1,ind-2} fera prochainement l'objet d'un hommage à la Société Asiatique de Paris \cite{JA}. On dégage ici la ligne directrice de ses travaux ; on suggère également la pertinence de ses conceptions pour la mesure du temps propre dans les systèmes non-inertiels.

Dans chacun de ses travaux, sa clarté d'analyse, sa virtuosité technique et sa passion pour la Science sont évidentes. Ceux qui l'ont croisé pourront attester, qui de son intelligence, qui de sa générosité, qui de sa bienveillance, qui de son désintéressement. Il est de ces rares auteurs dont le style reflète une conception de la Science libérée de toute considération personnelle. Son travail se place dans la continuité de la tradition de la Physique moderne telle qu'elle était pratiquée dans le laboratoire de Physique Théorique de l'Institut Henri Poincaré, sous l'égide de Louis de Broglie \cite{db}. Du point de vue mathématique, ses conceptions ont été formées notamment par la lecture d'\'E. Cartan et de Lichnerowicz. Sa conception de la notion de variété est issue de celle qu'explorait Cartan dans les années vingt, lorsqu'il cherchait à préciser la notion d'observateur en Relativité Générale. En termes imagés, S. Kichenassamy a montré que, s'il est naturel d'admettre que différents  observateurs puissent repérer les mêmes événements, l'expérience montre qu'ils n'ont pas la même grille de lecture pour interpréter leurs mesures les concernant, notamment lorsqu'ils sont soumis à un champ de gravitation.

Son \oe uvre peut être vue comme le déploiement d'un raisonnement développé continûment depuis ses premiers travaux, visant à mettre au jour ce que, à la lumière de la Physique Moderne, un observateur peut effectivement accomplir---sans supposer \emph{a priori} l'existence d'un observateur privilégié dont l'interprétation s'imposerait aux autres. Nous nous efforcerons ici de dégager la logique des arguments les plus novateurs, en supposant présents à l'esprit du lecteur le concept de système d'inertie, et les principes d'équivalence forte et faible.  Des éléments de cette analyse se trouvent dans ses synthèses de ses résultats \cite{matscience,GTPM,interpretation,CGR,apercu-anciens}, et dans ses cours et textes inédits.

La fa\c con dont il convient d'aborder ses travaux peut être illustrée en rapprochant trois de ses articles. D'une part celui où il clarifie nombre de questions autour du \og principe de Palatini \fg\ dans les théories lagrangiennes de la gravitation \cite{palatini} : il y montre que l'emploi systématique de multiplicateurs de Lagrange associés \`a des contraintes géométriques permet de réconciler différentes méthodes de variation proposées antérieurement, et de clarifier la structure de ces théories, tout en jetant une lumière nouvelle sur leur interprétation physique. Les deux autres sont des articles historiques. L'un montre comment l'évolution du calcul des variations avait conduit à introduire plusieurs types de variation fonctionnelle dont l'une---celle qui s'interprète souvent comme une dérivée de G\^ateaux---a fini par \^etre privilégiée \cite[(b)]{hist}. L'autre retrace l'évolution de la théorie des spineurs sur une variété \cite[(a)]{hist}. En lisant ces trois travaux, il est apparent que les deux derniers fournissent la justification des choix opérés dans le premier. Mais il est difficile d'inférer à partir du premier d'entre eux la richesse des arguments développés dans les deux autres. La plupart de ses travaux présupposent ainsi des bases théoriques, dont il a révélé les implications nouvelles par leur intégration dans un développement cohérent. La cohérence est le maître mot de son \oe uvre: ses travaux se fondent sur une intégration complète de tous les résultats antérieurs. C'est pourquoi son point de vue est souvent le seul compatible avec ce que l'on sait à l'heure actuelle.

Après quelques repères biographiques, nous suivrons la logique de son raisonnement, qui reprend les concepts de temps et d'espace à la base, décrivant au fur et à mesure ses conséquences vérifiables, auxquelles il a donné la première place dans tous ses travaux.

\section{Repères biographiques}

S. Kichenassamy \cite{nom} est né à Bahour (Pondichéry) le 22 juin 1926. Sa scolarité fut extr\^emement brillante : il réussissait tout sans effort. Autant que son intelligence, son intégrité et ses engagements lui valurent très tôt l'estime de tous. Il restera une référence, tant scientifique que morale. Il semble qu'il ait réalisé ce que son père pressentait qu'il devait accomplir.

Il voit très tôt en la Physique Théorique moderne une analyse des conditions de validité de tout discours scientifique qui se fonde sur la mesure, et décide de s'y consacrer. Cette démarche s'inscrit dans le cadre d'une interrogation sur la nature de la réalité, dans le droit fil de certaines écoles philosophiques tamoules.

Les circonstances font qu'il n'arrive à Paris qu'en janvier 1947, pour des études supérieures de Physique. L'environnement scientifique était exceptionnel ; les conditions matérielles, spartiates, dans la France des tickets de rationnement, plus encore pour un végétarien, loin de sa famille, qui de surcroît n'avait jamais connu le froid. Les uns après les autres, ses camarades pondichériens repartaient. Il est resté. Il fut vite remarqué, tant par les physiciens que par les indianistes \cite{tamoul}. Il devint le premier Pondichérien Docteur ès Sciences Mathématiques \cite{coord}. Il fera toute sa carrière à Paris, au sein du groupe de Louis de Broglie, dans le laboratoire de Physique Théorique, situé d'abord à l'Institut Henri Poincaré, puis à Jussieu. En 1962, alors qu'il avait prévu un voyage en Inde, il re\c cut une invitation non sollicitée de Princeton. Il choisit l'Inde et, de manière imprévue, se maria. Ma mère décède en 1969, à l'\^age de trente-quatre ans. Il a depuis continué son \oe uvre tout en assumant pleinement la charge d'élever seul son fils. Ce n'est pas le lieu de développer son impact sur chacun des travaux de ce dernier.

En tant que chercheur, son enseignement était exclusivement de niveau \og troisième cycle \fg. Fait rarissime, il lui est arrivé d'\^etre applaudi à la fin de ses cours. Directement ou indirectement, il a marqué tous les relativistes fran\c cais, particulièrement ceux qui ont été formés à cette époque. Au début des années soixante, il est invité plusieurs fois au Brésil, à l'Université du Paran\'a (à Curitiba), pour aider à la mise en place d'un programme de recherches en Physique Théorique. Il a gardé un vif souvenir de la chaleur de l'accueil de tous les membres de l'Université, des étudiants aux plus hautes autorités. Il fut l'un des acteurs de la réorganisation de l'enseignement de Physique Théorique à Paris \cite{GTPM}. Membre du Conseil Scientifique de l'Université Paris VI lors de sa création, il fut également l'un des deux membres du Collège A qui ont permis d'y maintenir la Physique Théorique (l'autre était O. Costa de Beauregard).

On dénombre une soixantaine publications en Physique Théorique---avec souvent une forte composante mathématique---la plupart dans des revues internationales, couvrant tout le spectre de la Relativité, Restreinte et Générale. Il faut y ajouter plusieurs cours et textes inédits. Il a également animé un groupe très actif dont quelques résultats marquants sont résumés dans \cite{CGR}. Qu'il s'agisse d'un article ou du texte d'une conférence, chaque travail contient généralement des arguments originaux qui ne sont pas repris dans ses autres publications (voir par exemple \cite{matscience,interpretation,CGR,rev}).

\section{De la crise de la Théorie Quantique des Champs à la critique des concepts d'espace et de temps}

\subsection{Interaction matière-rayonnement}

Mon père remarqua dans la thèse de B.G. Gokhale (qui sera publiée en 1952 \cite{gokhale}) un résultat surprenant : une loi en $Z^3$ des largeurs des raies des éléments de ${}_{37}$R à ${}_{50}$Sn. Il a pu en rendre compte, au moyen des transitions Auger \cite{mq}. Par ailleurs, gr\^ace à une interprétation du terme quadratique en potentiel vecteur de l'expression non-relativiste de l'hamiltonien d'interaction du système (électron $+$ rayonnement), il a pu expliquer des résultats expérimentaux de Lennuier au voisinage de la résonance optique  \cite{mq}. Pour le \og patron\fg\ (Louis de Broglie) il s'agissait là de \og vraie physique \fg, o\`u les considérations théoriques permettaient une meilleure compréhension des phénomènes \cite{ldb}. Ce sera une constante de son \oe uvre.

L'analyse des nombreux travaux sur lesquels on sollicitait son avis \cite{rapports}, a suscité chez lui dès les années cinquante une saine méfiance envers tout raisonnement qui n'explicite pas ses présupposés et pense pouvoir faire l'économie de la critique des concepts sur la base de notions de \og sens commun \fg\ qui, soumises à l'examen, s'avèrent préscientifiques \cite{proust}. Il fut ainsi conduit à une exploration de plus en plus profonde des concepts fondamentaux de la Physique.

\subsection{Place centrale des considérations relativistes}

C'était l'époque de la crise de la Théorie Quantique des Champs. Chacun voyait que les notions fondamentales de la mécanique quantique n'étaient pas compatibles avec la Relativité. En effet, en Relativité Restreinte, les notions de corps rigide, de rotation uniforme, de centre de masse conduisent à des incohérences : presque tout ce qui, en Mécanique newtonienne, fait intervenir plusieurs observateurs partageant un espace commun, cesse d'avoir un sens bien déterminé. La signification des concepts fondamentaux n'est pas claire, même si l'on arrive à \og sauver les phénomènes\fg\ (\emph{s{\=o}zein ta phainomena}). La situation est résumée par Wigner \cite[p.~263]{wigner-57} :
\begin{quote}
It remains true that we consider, in ordinary quantum theory, position operators as observables without specifying what the coordinates mean. [...] If we analyze the way in which we ``get away'' with the use of an absolute space concept, we simply find that we do not. In our experiments, [...] the so-called observables of the microscopic system are not only not observed, they do not even appear to be meaningful.
\end{quote}
La situation est d'autant plus grave que beaucoup de vérifications expérimentales de la Relativité Générale font intervenir des phénomènes essentiellement quantiques ; on en trouvera de nombreux exemples dans les travaux de mon père. Il y a donc une solidarité entre les deux domaines, tout autant qu'une incompatibilité théorique dans l'état actuel des théories.

Il faut donc avant tout expliciter le sens des mesures de temps et d'espace effectuées par des observateurs macroscopiques, ainsi que la signification physique des systèmes de coordonnées. Or, l'interprétation de celles-ci repose sur une fiction, aussi intimement liée à la pratique que les objets mathématiques sur lesquels reposent les calculs, mais de nature différente. Par exemple, le problème de la durée de vie du méson est posé en termes de deux systèmes, dont l'un est le système propre de celui-ci, système d'inertie (de la Relativité Restreinte) qui serait attaché à cette particule. On suppose donc littéralement qu'on peut imaginer un système d'observateurs, voyageant avec le méson pour ainsi dire, munis d'appareils de mesure \og idéaux\fg, et qui ont mis leurs horloges à l'heure avec des signaux lumineux. Tout ceci n'est jamais mis en oeuvre effectivement, et ne pourrait l'être. Pourtant, cette construction est indispensable pour donner un sens physique aux calculs les plus élémentaires (voir \cite{temps} pour une discussion détaillée des postulats sous-jacents). La nécessité de telles constructions a déjà été soulignée par Einstein:
\begin{quote}
The prejudice---which has by no means died out in the meantime---consists in the faith that facts by themselves can and should yield scientific knowledge without free conceptual construction. Such a misconception is possible only because one does not become easily aware of the free choice of such concepts, which, through verification and long usage, appear to be immediately connected with the empirical material \cite{autobio}.
\end{quote}
L'enjeu est ici de préciser le concept de système d'inertie et de déterminer son domaine de validité : dans quelle mesure les systèmes attachés à des observateurs locaux, éventuellement soumis à un champ de gravitation, peuvent-ils être assimilés à des systèmes d'inertie ?

\section{Systèmes d'inertie de la Relativité Restreinte et réciprocité}

C'est dans les années soixante que les horloges atomiques remplacent définitivement les mesures astronomiques dans la définition de la seconde \cite{temps-atomique}. Supposons, ce qu'on admet généralement, que les horloges atomiques réalisent les \og horloges idéales \fg\ qu'Einstein envisageait et qui, en l'absence de champ de gravitation ou d'accélération, sont identiques à tous égards. Il faut alors, sur cette base, déterminer comment deux observateurs inertiels peuvent coordonner leurs repérages des événements, en tenant compte de la constance de la vitesse de la lumière, incompatible avec le groupe de Galilée. On obtient ainsi la transformation de Lorentz. Une seule dérivation de celle-ci obtient, sans la postuler \emph{a priori}, la réciprocité entre des systèmes d'inertie qui peuvent communiquer par des signaux électromagnétiques supposés à propagation isotrope à vitesse $c$ \cite{lorentz}. Résumons ce travail, qui représente une étape essentielle du raisonnement de cet article ; pour la démonstration complète, voir  \cite{lorentz}.

Considérons deux systèmes d'inertie $S$ et $S'$.\footnote{Les indices grecs vont de 0 à 3, les indices latins de 1 à 3, et $\eta_{\alpha\beta}$ représente la métrique de Minkowski.} Le système $S'$ est caractérisé par sa vitesse de translation uniforme $c\beta^i$ par rapport à $S$ ; de même, $S$ a la vitesse (uniforme) $c\beta^{i'}$ par rapport à $S'$ . La transformation de $S$ à $S'$ est donc une fonction du point et de $c\beta^i$. Cette transformation est affine et conforme (Weyl, Fock). Avec des notations évidentes, on a alors
\[ x^{\alpha'}=a^{\alpha'}_\sigma x^\sigma + b^{\alpha'} ;\quad
x^{\alpha}=a^{\alpha}_{\sigma'} x^{\sigma'} + b^{\alpha},\text{ avec }
a^\rho_{\alpha'}a^{\alpha'}_\sigma=\delta^\rho_\sigma ;
\]
de plus, le facteur conforme de la transformation de $S$ à $S'$ ne dépend que de $\beta^i$ : $\Lambda=\Lambda(\beta^i)>0$.
L'élément nouveau est le suivant : sans nouvelle hypothèse, la vitesse de $S$ par rapport à $S'$ est égale en grandeur à celle de $S'$ par rapport à $S$, et cette relation de réciprocité implique que $\Lambda=1$.

Les hypothèses physiques entraînent donc la réciprocité, laquelle montre que le facteur conforme vaut un. Or, il existe des transformations conformes pour lesquelles le facteur conforme possède une autre valeur \cite{necessaire,matscience}. Il y a donc deux classes de phénomènes : ceux pour l'étude desquels on peut considérer que les appareils de mesure idéaux coïncident avec ceux des observateurs réels, et alors, on peut utiliser les transformations de Lorentz, et ceux qui affectent même les appareils de mesure \og idéaux\fg. Pour ces derniers, la réciprocité ne va plus de soi. L'expérience suggère que les phénomènes affectés par un champ de gravitation appartiennent à la seconde classe.

Avant de quitter le domaine de la Relativité Restreinte, examinons les conséquences physiques de ce qui précède pour la première classe de phénomènes, au-delà des conséquences bien connues de la transformation de Lorentz.

\section{Conséquences en Relativité Restreinte}

Si l'on conçoit la transformation de Lorentz comme une correspondance entre deux observateurs inertiels, qui ne communiquent que par des signaux élec\-tro\-ma\-gnétiques, et non comme une transformation interne à un espace-temps de Minkowski, certaines questions classiques prennent un nouvel aspect. Ainsi, bien que chaque observateur inertiel définisse une décomposition canonique de son espace de Minkowski en temps et espace, ses axes d'espace ne sont déterminés qu'à une rotation spatiale près. Comme la composée de deux transformations spéciales de Lorentz diffère d'une transformation spéciale par une rotation spatiale, il s'ensuit que si l'on regarde les observateurs inertiels associés aux quadrivecteurs unitaires $u^\alpha$ tangents à la ligne d'univers d'une particule accélérée en chacun des points de sa trajectoire, ces observateurs ne peuvent pas identifier leurs axes d'espace de manière cohérente. Considérons en effet trois systèmes $S_1$, $S_2$, et $S_3$, correspondant à trois points consécutifs $M_1$, $M_2$ et $M_3$ de la trajectoire, et les transformations spéciales $L_{12}$, $L_{23}$ et $L_{13}$ (bien déterminées) reliant $S_1$ à $S_2$, $S_2$ à $S_3$ et $S_1$ à $S_3$ ; on a alors en général $L_{23}L_{12}\neq L_{13}$. Donc, même en Relativité Restreinte, il n'est pas possible de donner aux systèmes d'inertie une extension globale lorsqu'il s'agit de systèmes instantanément liés : on ne peut pas identifier de manière canonique l'espace de $M_1$ à l'espace de $M_3$, car on trouve des résultats différents selon que l'on passe par $M_2$ ou pas. Un système tel que $S_1$ est donc strictement local en général. La précession de Thomas s'interprète dans cet esprit sous une forme très simple \cite[(b)]{apercu-anciens}, sans introduire d'objets mathématiques autres que les tétrades.

Voici quelques autres situations concrètes où ce point de vue semble s'imposer ; on peut également les rattacher en dernière analyse à la relativité de la simultanéité à distance \cite[(b)]{apercu-anciens}.

\subsection{Formation des images}

Comme la vitesse de la lumière est finie, il faut distinguer cartographie et observation au moyen d'un signal physique. La forme d'un objet peut subir des distorsions qui ne se réduisent pas à la \og contraction de Lorentz\fg, mais qui s'interprètent en termes de la géométrie des sections d'espace de deux observateurs. On peut ainsi distinguer les différents modes de formation d'une image photographique en Relativité Restreinte \cite{image}.  Dans le même ordre d'idées, la séparation angulaire de deux objets $A$ et $B$, dans un univers cosmologique, peut être définie de manière cohérente, ce qui pertinent dans le problème de l'éjection relativiste (où $A$ éjecte $B$) \cite{ejection}.

\subsection{Déplacement rigide d'une surface}

Le problème du déplacement rigide d'une couche superficielle trouve une solution plus satisfaisante que le problème du corps rigide: l'expansion et le cisaillement de la congruence des lignes d'univers se compensent, ce qui explique que le déplacement rigide soit possible \cite{rigidity}. H. Devaux remarquait, à propos des lames monomoléculaires d'albumine \cite{devaux} :
\og J'avais reconnu dès cette époque [en 1903], non sans surprise \cite{mere} qu'une goutte d'albumine déposée sur l'eau s'étend à la manière de l'huile, prend des dimensions finales de m\^eme ordre de grandeur, et qu'elle subit en m\^eme temps une véritable coagulation, car elle devient une lame manifestement rigide.\fg\ Faut-il voir dans les lames monomoléculaires des modèles  d'objets susceptibles de déplacements rigides compatibles, avec une bonne approximation, avec la Relativité Restreinte ?

\subsection{Rotation relativiste}

Il n'est pas cohérent de décrire la vitesse linéaire d'un corps en mouvement circulaire uniforme en Relativité Restreinte par une relation de la forme $v = \omega r$, car cette vitesse peut prendre des valeurs plus grandes que $c$. On dispose néanmoins d'une transformation relativiste, celle de Franklin-Trocheris-Takeno (proposée en 1922 \cite{franklin}, redécouverte plusieurs fois par différentes méthodes), pour laquelle $v/c=\tanh(\omega r/c)$ \cite[(c)]{rotation}. Il est essentiel de la considérer comme une transformation \og à $r$ et $z$ constants \fg, sous peine d'aboutir à des incohérences \cite[(c), p.~5728]{rotation}. En effet, le mouvement s'effectue sous ces contraintes. Les systèmes correspondant aux rayons $r$ et $r+dr$ sont distincts, et sont précisément reliés par une transformation de Lorentz si l'on adopte la transformation de Franklin-Trocheris-Takeno \cite{rotation,apercu-anciens} ; toute autre transformation serait en conflit avec la loi relativiste de composition des vitesses.

Dans la description des pulsars, la rotation relativiste permet d'éviter la notion de \og cylindre de lumière\fg, lieu fictif défini par une relation de la forme $\omega r=c$, au-delà duquel les équations de Maxwell changeraient de type (sic). On rétablit ainsi la covariance des équations de l'électromagnétisme \cite[(c)]{rotation}. On trouve également une confirmation de ce point de vue dans l'étude du mouvement d'une particule chargée. La trajectoire spatio-temporelle de la particule est décrite de manière intrinsèque par trois quantités \cite{FS} ; on montre que l'on peut les rattacher, dans le cas du champ constant, aux invariants du champ et à la position de la direction de la particule par rapport \`a deux 2-plans intrinsèquement attachés au champ. Pour ces résultats et une étude de la généralisation du mouvement hélicoïdal, voir \cite{charged}.

\subsection{Section efficace}

Parmi les questions de cinématique relativiste, mentionnons le problème de la définition de la section efficace de diffusion Compton \cite{section-efficace}. Les mesures de l'interaction entre matière et rayonnement s'effectuent souvent en termes de coefficients de diffusion ou d'affaiblissement et non de sections efficaces, comme on le voit dans les traités classiques \cite{coeff_diff}. Les sections efficaces données dans les tables récentes sont souvent déduites par une simple convention mathématique de ces coefficients. La notion de section efficace n'intervient pas, par exemple, dans le travail original de Klein et Nishina, ni dans sa vérification expérimentale par Meitner et Hupfeld \cite{kn-mh}. Mais dans d'autres travaux, on introduit une description en termes de sections efficaces, en rupture par rapport à ces travaux classiques. La question de la variance relativiste des quantités en cause se pose alors. On considère dans cette section les définitions de la section efficace ; les coefficients classiques seront évoqués dans la section suivante, à propos des théories cinétiques relativistes.

S'il y a $d\nu$ collisions entre deux espèces de particules, de densités numériques $n_1$ et $n_2$ dans le quadrivolume $dVdt$, on définit la section efficace $\sigma$ par
\[ d\nu = \sigma v_{12}n_1n_2dVdt ;
\]
la question est de s'accorder sur une convention sur la valeur de la vitesse relative $v_{12}$, sachant que la combinaison
\[ \sigma v_{12}(1- \mathbf v_1\cdot\mathbf v_2)^{-1}
\]
est invariante. Si l'on prend pour $v_{12}$ la vitesse relative habituelle $v_R$, définie par
\[ v_{R}^2=\frac{{|\mathbf v_1-\mathbf v_2|^2-|\mathbf v_1\wedge\mathbf v_2|^2}}{(1-\mathbf v_1\cdot\mathbf v_2)^2},
\]
on trouve que $\sigma(1- \mathbf v_1\cdot\mathbf v_2)^{-1}$ est un invariant. Si l'on choisit de poser que $\sigma$ est invariante, $v_{12}$ ne l'est plus ; pour le cas de la lumière, on obtient ainsi le résultat choquant que la vitesse de la lumière n'est plus égale à $c$ dans tous les systèmes d'inertie (sic). Pour les autres incohérences auxquelles conduit ce choix, voir \cite[(b), pp.~252-253]{apercu-anciens}, \cite{section-efficace}. Heureusement, le traitement relativiste correct basé sur la vitesse relative invariante est très simple, et permet de justifier l'introduction, qui sinon serait purement empirique, d'un facteur supplémentaire dans le calcul du flux total \cite[\'eq.~(2)]{compton-n}.

Considérons maintenant les théories cinétiques relativistes, qui mettent en cause, entre autres, les quantités effectivement mesurées tels que les coefficients d'absorption \cite{fluids}. Ces travaux ont été préparés par un cours d'introduction aux Théories Cinétiques Classiques à Paris en 1974--1975.

\subsection{Théories cinétiques et transfert de rayonnement}

La théorie cinétique classique repose sur la distribution maxwellienne des vitesses ; celle-ci autorise des vitesses arbitrairement grandes et de plus, elle est intimement liée à l'addition non-relativiste des vitesses. On dispose d'une maxwellienne relativiste, la distribution de Jüttner-Synge, mais la modification de la maxwellienne n'est pas tout : (i) les conditions de normalisation ne sont pas toujours invariantes \cite[éq.~(4)]{diffuse} ; (ii) les lois de transformation des quantités observables ne sont pas connues \emph{a priori} ; (iii) le passage à la limite non-relativiste ne se réduit pas à une approximation \og aux vitesses faibles \fg\ \cite[éq.~(14)]{diffuse}. Outre une discussion de ces points, on trouvera dans \cite[p.~185]{diffuse} les expressions de l'intensité $I$, du coefficient d'émission $\ep$ et du coefficient de diffusion $\sigma$ en termes de la fréquence du photon et des quantités invariantes.

Cette remise à plat a permis de proposer une équation de Liouville relativiste \cite{liouville}, d'aborder le problème de la diffusion par un gaz en translation uniforme \cite{diffuse}, de donner une nouvelle forme (i) des relations d'Einstein-Milne \cite{einstein-milne}, (ii) de la fonction de redistribution pour la diffusion de la lumière par des électrons relativistes \cite{redistribution}, ainsi que (iii) de l'équation de Saha \cite{saha}, avec une discussion quantitative du domaine de validité de ces résultats. En outre, on peut formuler une équation de transfert radiatif en Relativité Générale pour un milieu isotrope dispersif, faiblement absorbant, non-magnétique \cite{dispersive}. Dans le cas de l'élargissement Doppler des raies d'absorption, on obtient une nouvelle contribution \cite{doppler-broadening}, m\^eme à l'approximation non-relativiste, à cause de la non-invariance Galiléenne du coefficient d'absorption volumique $\chi_\nu$. Comme la propagation de la lumière n'est pas invariante par le groupe de Galilée, ce résultat n'aurait pas pu \^etre obtenu en Physique pré-relativiste \cite[p.~655]{doppler-broadening}. La définition des coefficients est obtenue en les interprétant comme des coefficients dans une équation de transfert radiatif invariante, ce qui fournit la variance des quantités observées \cite{einstein-milne}.
Certains de ces résultats sont d'ailleurs établis en Relativité Générale, voir par exemple \cite{dispersive}.

Ceci nous amène à l'étape suivante du raisonnement : si le mouvement uniforme et l'électromagnétisme ne peuvent être décrits de manière satisfaisante que dans le cadre de la Relativité Restreinte, qu'en est-il des mouvements accélérés ? Quelle est la représentation adaptée à la description d'expériences effectuées dans un système non inertiel ?

\section{Le problème du système entraîné et la C-équiva\-lence}

En mécanique newtonienne, les mouvements s'effectuent dans une arène (l'espace absolu), et tous les systèmes d'inertie sont équivalents par le groupe de Galilée. Il suffit de déterminer (approximativement) des coordonnées cartésiennes de cet espace pour rapporter toutes les mesures à un jeu de coordonnées essentiellement unique qui mettrait en évidence les symétries de l'espace. L'abandon de la mécanique newtonienne ne signifie pas que les systèmes de coordonnées privilégiés n'existent pas, mais que leur construction repose sur des hypothèses dont la signification doit être explicitée \cite{cond-coord}. La Relativité Restreinte, dont l'espace-temps de Minkowski pourrait être vu comme une arène, semble encore admettre des systèmes de coordonnées privilégiées adaptées à des groupes de symétrie. C'est l'étude de la Relativité Générale et des théories unitaires \cite{e-s} qui a d'abord conduit mon père à mettre au jour une difficulté dans ce point de vue global : \og la manière dont [les] champs [à symétrie sphérique ou axiale] sont déterminés est peu convaincante ; on s'aper\c coit notamment que les différentes symétries sont définies non dans l'espace-temps, mais dans un espace euclidien auxiliaire rapporté aux m\^emes coordonnées\fg\
\cite[(c), p.~7-02]{coord}. Pour échapper à cette difficulté, il a introduit dans sa Thèse \cite{coord}, notamment sur la base du traitement de la dérivée de Lie dans le traité de Lichnerowicz sur les groupes de transformations \cite{groupes}---seul traité moderne à sa disposition---sa méthode des transformées infinitésimales pour l'étude des solutions exactes. Il a également montré peu après comment obtenir les identités de conservation en théorie unitaire sans faire appel aux équations de liaison, ce qui clarifie plusieurs travaux antérieurs \cite{conservation}.

Si l'on se place au contraire dans un espace sans symétrie, certaines des lois de conservation sont problématiques : ainsi, la conservation de l'énergie résulte de la conservation du tenseur impulsion-énergie combinée à l'existence d'un vecteur de Killing orienté dans le temps \cite[(d)]{acc}. La situation est différente dans le cadre cosmologique, où l'on postule que l'univers dans son ensemble peut être décrit à grande échelle par une seule variété, ce qui peut conduire à se donner, entre autres, une variable temps globale. Mais un observateur local n'a pas nécessairement accès aux mêmes informations qu'un tel observateur global. Peut-on postuler que les unités d'un observateur terrestre réel coïncident avec celles d'un observateur global putatif ?

L'interprétation physique des mesures d'espace et de temps est donc profondément modifiée en l'absence d'informations semi-globales telles que l'existence de symétries, ou d'un système d'inertie étendu.

\subsection{Introduction de la C-équiva\-lence}

La Relativité Générale est une tentative pour sortir du cadre du système d'inertie global. L'idée de départ est bien connue : si l'on met un avion ou un \og ascenseur\fg\ sur une trajectoire de chute libre, il devrait se trouver en situation de \og gravité zéro\fg\ : y a-t-il donc dans certains cas équivalence entre un champ de gravitation et un champ d'accélérations ? Si oui, cela ne permet-il pas de d'affirmer que la Relativité Restreinte est localement valable, même si elle ne l'est pas globalement ?

Il faut distinguer équivalence faible et équivalence forte. L'équivalence faible est avant tout un principe qui stipule que les trajectoires d'une particule d'épreuve dans un champ de gravitation sont indépendantes de sa masse. Donc, s'il existe une variété fondamentale dans laquelle on repère les événements, la trajectoire d'une telle particule devrait être une courbe privilégiée de cette variété, une géodésique par exemple. En fait, cette propriété a été prise comme postulat (on sait qu'il est surabondant, voir par exemple \cite{eq-mvt}). On est ainsi conduit à postuler que la variété fondamentale est munie d'une connexion symétrique. En poursuivant ce raisonnement, on peut montrer que la Relativité Générale est une \og extension nécessaire de la Relativité Restreinte\fg\ \cite{necessaire}.

Peut-on alors identifier les observateurs locaux aux espaces tangents minkowskiens de la variété des événements ? Obtient-on ainsi des systèmes d'inertie locaux ? Si c'était le cas, tous ces espaces devraient par définition posséder les mêmes appareils de mesure \og idéaux\fg, donc les mêmes unités de mesure, et l'on ne voit pas comment les décalages spectraux gravitationnels pourraient s'expliquer : si le champ de gravitation affecte, par exemple, les atomes d'une horloge atomique, le postulat fondamental de l'identité de toutes les horloges idéales est en défaut ; s'il ne les affecte pas, il n'y a pas de décalage gravitationnel. Si les horloges idéales sont des objets matériels, il faut conclure que la Relativité Générale est incomplète. L'échec des nombreuses théories minkowskiennes de la gravitation et des théories unitaires nous conforte dans l'idée que la Relativité Générale doit pourtant former la base de la suite du raisonnement : avant de chercher à modifier la Relativité Générale, il faut s'efforcer d'amender son interprétation physique.

Le problème est de savoir ce qu'on entend par système propre pour une particule accélérée, et de décider si l'on le considère ou non comme un système d'inertie (dans lequel la Relativité Restreinte serait valable). Il faut distinguer deux notions : le \emph{système instantanément lié}, dont la vitesse en un point est celle du mobile, mais dont l'accélération est nulle, et le \emph{système entraîné}, dont l'accélération n'est pas nulle. Si l'on suppose que l'on peut assimiler le système entraîné à un système d'inertie, cela signifie que l'on fait les hypothèses suivantes : (a) les descriptions locales des phénomènes sont indépendantes du domaine de champ de gravitation considéré ; (b) ces descriptions locales sont toutes identiques à celle que l'on aurait en Relativité Restreinte, tant du point de vue des relations mutuelles que des valeurs absolues, ce qui revient à supposer (b1) l'isotropie locale et (b2) l'existence d'une jauge absolue, les appareils de mesure standard déterminant dans ces domaines la même jauge que par rapport à un sytème d'inertie. L'équivalence faible et l'isotropie locale semblent bien vérifiées par l'expérience \cite[\S~19]{necessaire} : on peut donc admettre (a) et (b1). Par contre, l'équivalence forte stipule l'existence au moins théorique d'appareils de mesure ayant partout le comportement minkowskien, bien que par ailleurs on suppose que le champ de gravitation influence les phénomènes physiques et les appareils de mesure effectivement utilisés : comme les phénomènes physiques semblent bien subir l'influence des champs de gravitation, on est conduit à remplacer le postulat (b2) par la \textbf{C-équivalence} : \emph{Tous les observateurs sont pourvus d'appareils de mesure standard (i.e., des appareils ayant même comportement au même point et au repos relatif). Mais, en raison du champ de champ de gravitation, ces appareils ont un comportement variable d'un point à l'autre de telle sorte que, pour deux événements séparés par des intervalles de coordonnées $\Delta x^a$, les appareils de mesure en $A$ donneront la valeur
\begin{equation}\label{eq:C-eq}
    (ds^2)_A=\Lambda_A\overline{ds}^2,
\end{equation}
où $(ds^2)_A$ représente l'intervalle spatio-temporel carré séparant $A(x^a)$ du point $B(x^a+dx^a)$, suivant un observateur situé en $A$, et  $\overline{ds}^2$ l'intervalle carré que leur attribuerait un observateur inertiel.}

Cette théorie s'est cristallisée dans un article soumis le 1\ier\ octobre 1963 \cite{complements}, et résumée dans une note que L. de Broglie présentera le 13 janvier 1964 \cite{fond-GR}. Si elle se substitue à l'équivalence forte, la C-équivalence est compatible avec le principe d'équivalence faible \cite{complements,redshift}. \og Ainsi, l'hypothèse de l'équivalence faible et l'adoption résultante d'étalons locaux (ceux du système d'inertie modifiés par le champ de gravitation) distincts d'un point à une autre de l'espace-temps de la Relativité générale, permettent de rendre compte du décalage vers le rouge des raies spectrales de manière intrinsèque et de distinguer, dans l'effet global, l'effet gravitationnel de l'effet Doppler\fg\ \cite[pp.~143--144]{complements}.

\subsection{La communication entre observateurs éloignés se fait à l'aide de champs singuliers}

Si deux observateurs sont réellement indépendants, ils peuvent ne pas être en mesure de comparer directement leurs mesures autrement que par échange de signaux électromagnétiques, et c'est cette analyse qui conduit à l'explication du décalage spectral gravitationnel. Il semble que, si l'importance des signaux lumineux est déjà visible dans le procédé de mise à l'heure,  l'étude de la polarisation de la lumière monochromatique dans un espace Lorentzien n'avait pas été approfondie. Il s'agissait d'étudier les champs électromagnétiques singuliers (par définition, ce sont ceux dont les deux invariants sont nuls). On montre que pour les champs singuliers généraux, \og le champ de gravitation agit à l'égard de la polarisation comme un milieu polarisable \fg\ \cite[(a), p.~267]{null}. L'étude complète de la variation des paramètres généralisés de Stokes, dont on montre qu'elle est cohérente avec la C-équivalence, permet d'atteindre des prédictions accessibles à la mesure \cite[(c), p.~14]{null}. On précise au passage les résultats associés aux noms de Robinson et Mariot, et de Goldberg et Sachs. Ajoutons que le cas singulier des théories de Born-Infeld ne fournit rien de nouveau \cite{born-infeld}.

C'est dans ce cadre que s'inscrivent les travaux \cite{acc} sur les mouvements accélérés, qui proposent une solution au problème souvent discuté de la définition de l'accélération et de la suraccélération uniforme \cite{acc,charged}. Rappelons en effet que le champ de rayonnement de l'électron accéléré est représenté par un champ singulier \cite[(c), p.~5]{null}.

On peut préciser le rapport entre les facteurs conformes en des points distincts, dans plusieurs situations concrètes \cite{complements,CGR}, \cite[(d)]{acc}. En voici quelques-unes : (1) Dans le cas du décalage spectral gravitationnel, on dispose d'une théorie assez développée. (2) Si l'on assimile le \og pseudo-système d'inertie \fg\ local au développement du second ordre d'une métrique d'espace-temps, il est naturel de remplacer, sous l'hypothèse d'isotropie locale (compatible avec l'expérience \cite[\S~19]{necessaire}), les transformations de Lorentz par des transformations conformes, qui décriraient le passage à ce pseudo-système d'inertie dans un voisinage de l'origine sur lequel le champ de gravitation serait sensible. (3) Dans la mesure o\`u, dans les mouvements uniformément accélérés, on peut calculer la relation entre accélération et temps propre, il est naturel de postuler une relation entre les deux ; en termes concrets, certains accéléromètres pourraient fournir un nouveau mode de mesure directe du temps propre. (4) Pour le cas de l'énergie de Schott, on peut voir l'effet de la relaxation du facteur conforme sur le bilan d'énergie.

\section{Conclusion et perspectives}

La C-équivalence demeure la seule réponse cohérente au problème de l'interprétation physique de la Relativité Générale. Elle conduit à admettre une structure de variété dans laquelle on représente les événements et la propagation des signaux, et à distinguer celle-ci de la variété des observateurs par un facteur local. Ce facteur conforme n'est pas un champ scalaire sur une variété des événements, parce que les observateurs ne sont pas entièrement déterminés par leurs points-origine : ainsi, plusieurs observateurs de vitesse et d'accélération différentes pourraient passer par le même point-événement. La signification du terme \og local\fg\ doit toujours être précisée par référence à une mesure effectivement réalisable par des observateurs humains (macroscopiques par essence). Les postulats sous-jacents aux fictions mathématiques nécessaires aux calculs peuvent être explicités, comme nous l'avons vu. En retour, ce nouveau point de vue modifie notre regard sur la Relativité Restreinte, avec les conséquences que nous avons résumées.

En définitive, la critique de la Mécanique Newtonienne et de la Relativité Restreinte, tant du point de vue de la cinématique que de la dynamique, conduit à la conclusion que les observateurs locaux effectuent des mesures sur la variété des événements, mais ne peuvent être identifiés ni à des voisinages étendus de celle-ci, ni à ses espaces tangents (lesquels ne font pas référence aux lignes d'univers des observateurs). En termes concrets, tous les observateurs ont accès aux mêmes événements, mais ne leur attribuent pas nécessairement la même signification, si ce n'est par une hypothèse arbitraire. Ils utilisent en effet des appareils de mesure semblables mais non identiques.

Ainsi, les travaux que nous venons d'évoquer présentent une profonde unité ; ils proposent des défis nouveaux aussi bien aux Mathématiques qu'à la Physique Théorique et Expérimentale, et montrent, sur des exemples pertinents pour les applications, comment les relever.


\end{document}